\documentclass[12pt,preprint]{aastex}

\def\dex#1{\hbox{10$^{#1}$}}
\def\tdex#1{\hbox{$\times10^{#1}$}}
\def\deg{$^\circ$}
\def\cmm#1{\hbox{${\rm cm^{-#1}}$}}

\def\kms{km\,s$^{-1}$}
\def\mA{m\AA}
\def\e{$\pm$}
\def\Msun{M$_\odot$}
\def\Msunpyr{M$_\odot$\,yr$^{-1}$}
\def\HI{\protect\ion{H}{1}}
\def\Ha{H$\alpha$}

\def\CaII{\protect\ion{Ca}{2}}
\def\FeI{\protect\ion{Fe}{1}}
\def\FeII{\protect\ion{Fe}{2}}

\begin{document}
\title{Distances to Galactic high-velocity clouds. Complex~C}

\author{B.P. Wakker\altaffilmark{1}, D.G. York\altaffilmark{2}, C. Howk\altaffilmark{3}, J.C. Barentine\altaffilmark{4}, R. Wilhelm\altaffilmark{5}, R.F. Peletier\altaffilmark{6}}
\author{H. van Woerden\altaffilmark{6}, T.C. Beers\altaffilmark{7}, \v{Z}. Ivezi\'c\altaffilmark{8}, P. Richter\altaffilmark{9}, U.J. Schwarz\altaffilmark{6,10}}

\altaffiltext{1}{Department of Astronomy, University of Wisconsin, Madison, WI 53706; wakker@astro.wisc.edu}
\altaffiltext{2}{Astronomy \& Astrophysics Center, University of Chicago, Chicago, IL 60637; don@oddjob.uchicago.edu}
\altaffiltext{3}{Dept. of Physics, University of Notre Dame, Notre Dame, IN 46556; jhowk@nd.edu}
\altaffiltext{4}{Dept. of Astronomy, Univ. of Texas, Austin TX 78712 and Apache Point Observatory, Sunspot, NM 88349; jcb@astro.as.utexas.edu}
\altaffiltext{5}{Dept. of Physics \& Astronomy, Texas Tech University, Lubbock TX 79409; ron.wilhelm@ttu.edu}
\altaffiltext{6}{Kapteyn Astronomical Institute, University of Groningen, Postbus 800, 9700 AV, Groningen; peletier@astro.rug.nl, hugo@astro.rug.nl}
\altaffiltext{7}{Department of Physics and Astronomy, Michigan State University, CSCE: Center for the Study of Cosmic Evolution and JINA: Joint Institute for Nuclear Astrophysics, Michigan State University, E.\ Lansing, MI 48824, USA; beers@pa.msu.edu}
\altaffiltext{8}{Dept. of Astronomy, University of Washington, Box 351580, Seattle, WA 98195; ivezic@astro.washington.edu}
\altaffiltext{9}{Institut f\"ur Physik, Universit\"at Potsdam, Am Neuen Palais 10, 14469 Potsdam, Germany; prichter@astro.physik.uni-potsdam.de}
\altaffiltext{10}{Department of Astrophysics, Radboud University Nijmegen, Toernooiveld 1, NL-6525 ED, The Netherlands; schwarz@kun.nl}

%%%%%%%%%%%%%%%%%%%%%%%%%%%%%%%%%%%%%%%%%%%%%%%%%%%%%%%%%%%%%%%%%%%%%%%%%%%%%%%%

\begin{abstract}
We report the first determination of a distance bracket for the high-velocity
cloud (HVC) complex~C. Combined with previous measurements showing that this
cloud has a metallicity of 0.15 times solar, these results provide ample
evidence that complex~C traces the continuing accretion of intergalactic gas
falling onto the Milky Way. Accounting for both neutral and ionized hydrogen as
well as He, the distance bracket implies a mass of 3--14\tdex6~\Msun, and the
complex represents a mass inflow of 0.1--0.25~\Msunpyr. We base our distance
bracket on the detection of \CaII\ absorption in the spectrum of the blue
horizontal branch star SDSS\,J120404.78+623345.6, in combination with a
significant non-detection toward the BHB star BS\,16034-0114. These results set
a strong distance bracket of 3.7--11.2~kpc on the distance to complex~C. A more
weakly supported lower limit of 6.7~kpc may be derived from the spectrum of the
BHB star BS\,16079-0017.
\end{abstract}


%%%%%%%%%%%%%%%%%%%%%%%%%%%%%%%%%%%%%%%%%%%%%%%%%%%%%%%%%%%%%%%%%%%%%%%%%%%%%%%%

\section{Introduction}
\par The evolution of galaxies is strongly driven by the gas in the interstellar
medium. There is strong evidence for the infall of new material that provides
fuel for galaxy growth. This gas may originate in accreted satellite galaxies,
as gas tidally pulled out of passing galaxies, or from pristine intergalactic
gas. The cool, infalling clouds appear to be embedded in an extended
(100--200~kpc radius) hot Corona (Sembach et al.\ 2003). Indirect evidence for
infalling gas is provided by two arguments: (a) At the current rate of star
formation, all of the ISM will be turned into stars within about a Gyr. (b) The
narrowness of the distribution of metallicities of long-lived stars implies that
the metallicity of the ISM remains more or less constant over a Hubble time,
which can happen if there is a continuing inflow of low-metallicity material
with a present-day rate of about 1~\Msunpyr. Item (b) is known as the ``G-dwarf
problem'' (van den Bergh 1961). Using the infall hypothesis to solve it has been
the subject of much theoretical work (see e.g.\ Pagel 1997 for a good summary).
Continuing infall is essential in detailed numerical modeling of the chemical
evolution of the Galaxy and the development of abundance gradients (e.g.\
Chiappini et al.\ 2001 and references therein). Infall of low-metallicity gas
also seems necessary to reproduce the relatively high abundance of deuterium
measured in the local interstellar medium (Linksy et al.\ 2006).
\par Direct observational evidence for infalling low-metallicity gas is provided
by the high-velocity clouds (HVCs; see reviews by Wakker \& van Woerden 1997;
Richter 2006). Subsolar metallicities have now been determined for eleven clouds
(see van Woerden \& Wakker 2004 for a summary). In particular, the metallicity
of complex~C is well established as 0.15 times solar (see summary by Fox et al.\
2004). Complex~C also has a high deuterium abundance (Sembach et al.\ 2004).
Distance brackets have been more elusive, with just one known before 2006
(8--10~kpc for complex~A -- van Woerden et al.\ 1999a; Wakker et al.\ 2003).
Thom et al.\ (2006) derive an 8.8~kpc upper limit for cloud WW\,35, while in a
separate paper (paper~I, Wakker et al.\ 2007), we present new results for two
HVCs (9.8--15.1~kpc for complex~GCP and 5.0--11.7~kpc for the Cohen Stream). In
this letter we report a distance bracket for the HVC covering the largest sky
area -- complex~C. We summarize our method in Sect.~2. The data are described in
Sect.~3, the results in Sect.~4, while in Sect.~5 we summarize the implications.

%%%%%%%%%%%%%%%%%%%%%%%%%%%%%%%%%%%%%%%%%%%%%%%%%%%%%%%%%%%%%%%%%%%%%%%%%%%%%%%%

\section{Method}
\par To find the distance to a HVC, we search for interstellar absorption at the
cloud's velocity in spectra of stars with known distances. A detection sets an
upper limit, while a {\it significant} non-detection sets a lower limit. A
significant non-detection means that the ratio of the expected equivalent width
to the observed 3$\sigma$ upper limit is sufficiently large (e.g.$>$3; see
Appendix item 15 in Wakker 2001). We also refer to paper~I for a detailed
discussion.
\par We find probe stars from the HK survey (Beers et al.\ 1996), the Sloan
Digital Sky Survey (SDSS; Fukugita et al.\ 1996; Gunn et al.\ 1998, 2006; York
et al.\ 2000; Stoughton et al.\ 2002; Pier et al.\ 2003; Adelman-McCarthy et
al.\ 2007) and the 2MASS survey (Cutri et al.\ 2003; Brown et al.\ 2004). Using
color criteria we identify blue horizontal branch (BHB) and RR\,Lyrae
candidates, for which we then obtain intermediate-resolution spectra and
photometry to derive the stellar parameters. See paper~I for more details and
Wilhelm et al.\ (2008, in preparation) for a complete description.

%%%%%%%%%%%%%%%%%%%%%%%%%%%%%%%%%%%%%%%%%%%%%%%%%%%%%%%%%%%%%%%%%%%%%%%%%%%%%%%%

\section{Observations}
\subsection{WHT observations}
\par In 1997, we selected several stars with estimated distances between 2 and
8~kpc projected on the complex~C core CI (Fig~1). Five of these were observed by
Peletier and van Woerden with the 4.2-m William Herschel Telescope (WHT) on La
Palma on May 17/18/19 1997, using the Utrecht Echelle Spectrograph (UES) at the
WHT Nasmyth focus. The spectra covered the wavelength range from 3610 to
4510~\AA, with a velocity resolution of 6~\kms. The wavelength calibration was
obtained with a ThAr lamp. The standard IRAF reduction included bias and
flatfield corrections, summing of pixels across the spectrum, and sky
subtraction.
\par A preliminary analysis of these spectra and preliminary distances to the
stars yielded a lower limit of 6.1~kpc to the distance of complex~C (van Woerden
et al.\ 1999b; Wakker 2001). We have now derived final stellar distances for
three stars, using photometric data and intermediate-resolution spectra obtained
at the McDonald observatory 2.7m telescope. Table~1 presents the basic data for
these stars. All three are hot, high-gravity stars, and we carefully checked the
spectral details against the derived stellar parameters. For one star
(BS\,16086-0123) we do not have classification data, and one star
(BS\,16034-0002) turns out to be a cool, nearby, star.

\subsection{Keck observations}
\par In 2007, we selected several stars from the SDSS that lie projected onto
core CIII (see Fig.~1), with distances ranging from 9.3 to 40.6~kpc. These
distances are based on SDSS spectroscopic and photometric data. On April 24 2007
Barentine observed three of these using the upgraded High Resolution Echelle
Spectrometer (HIRES; Vogt et al.\ 1994) on the Keck~I telescope. The 2004 HIRES
upgrade replaced the single CCD with a three-CCD mosaic, including two with
enhanced blue sensitivity. All data were collected using the UV cross disperser
and the C5 decker (1\farcs15 wide slit). The data were binned by 2 pixels
(0\farcs24) in the spatial direction. The seeing was approximately 1\farcs5. 
The spectra cover the wavelength range 3380 to 4330~\AA\ and have a spectral
resolution of 8.8~\kms. Three stars could be observed before weather conditions
forced the telescope to close for the night. Two are useful (see Table~1), while
for SDSS\,J120557.16+625251.6 the stellar \CaII\ line obscures the interstellar
lines.
\par The spectra were extracted by Howk using the HIRedux package (v2.2) of
J.X.\ Prochaska.\footnote{Available through:
http://www.ucolick.org/$\sim$xavier/IDL/.} The two-dimensional echelle images
were bias-subtracted, flat-fielded, and wavelength-calibrated using the HIRES
ThAr and quartz (flat field) lamps. One-dimensional spectra are extracted using
an optimal extraction routine, and individual exposures and orders were co-added
with an inverse variance weighting. The blaze function of the instrument was
removed (before co-adding orders) by fitting a polynomial to the average flux
distribution of the orders within each of the three CCDs.

\subsection{\HI\ data}
\par We also have 21-cm \HI\ profiles toward the probe stars. Effelsberg spectra
(9\farcm7, 1~\kms\ resolution; Wakker et al.\ 2001) are available for the stars
observed with the WHT. For the Keck stars we use the LAB dataset of Kalberla et
al.\ (2005; 36\arcmin, 1~\kms\ resolution). However, Wakker et al.\ (2001) found
that $N$(\HI) measured with a 36\arcmin\ beam can be up to a factor 2.5 larger
or smaller than $N$(\HI) measured with a 9\farcm7 beam; the ratio distribution
has a mean of 1 and rms of 0.2. The \HI\ column densities therefore have a large
($>$20\%) systematic uncertainty. Higher resolution observations
($\sim$1\arcmin) with a synthesis telescope are needed to obtain more accurate
values.

\section{Results}
%\subsection{Upper limit to the distance of complex~C}
\par Columns 8--13 of Table~1 list the \HI\ and \CaII\ measurements, including
predictions for the equivalent width (EW) based on the relation between $N$(\HI)
and \CaII\ abundance found by Wakker \& Mathis (2000; see notes). Figure~2 shows
the \CaII~K and H and \HI\ 21-cm spectra for the four stars that yield
significant results.
\par We detect \CaII~K and H absorption associated with complex~C toward the
star SDSS\,J120404.78+623345.6, with EW(K)=42\e3\e4~\mA\ and EW(H)=19\e6\e3~\mA.
The first error is statistical error associated with the noise in the spectrum
and the placement of the continuum. The second error is a systematic error
associated with a 3~\kms\ uncertainty in choosing the velocity limits of the
equivalent width integration. See Wakker et al.\ (2003) for a full discussion of
these errors. That this line is interstellar is shown by two facts. (a) It is
much narrower (FWHM 6.7~\kms) than the stellar lines (FWHM $\sim$15~\kms). (b)
Stars with effective temperatures of about 7000~K do not have a stellar line at
this location (see Fig.~3). The distance of SDSS\,J120404.78+623345.6 is found
to be 10.9$\pm$0.7~kpc. As we discuss in Paper~I, this implies a 68\% confidence
interval for the upper limit on the distance of complex~C of
$D_*$+0.47$\sigma$($D_*$)= 11.2~kpc. 

%\subsection{Lower limit to the distance of complex~C}
\par The three WHT stars yield non-detections for \CaII\ in complex~C. The
spectrum of BS\,16079-0017 ($D$=8,1$\pm$2.9~~kpc) shows many broad stellar
lines, but no narrow interstellar line is visible at the velocity of complex~C.
This star thus sets a tentative lower limit of 6.7~kpc to the distance of
complex~C. On the other hand, a firm lower limit of 3.7~kpc is set by
BS\,16034-0114 ($D$=3.8$\pm$0.3~kpc), whose spectrum shows few stellar lines
and which has EW(expected)/3$\sigma$(EW)=11. The star BS\,16079-0015 also yields
a significant non-detection, but since it is closer than BS\.16034-0114, we do
not show its spectrum in Fig.~2.
\par Complex~C is also not detected toward the star SDSS\,J121611.13+645811.0,
even though this star is more distant than SDSS\,J120404.78+623345.6. However,
the expected EW is only a factor 2.5 higher than the 3$\sigma$ limit.
Considering possible intrinsic variations in the \CaII\ abundance, and the 
large uncertainty in the \HI\ column density (see above), this non-detection is
not considered significant, though only marginally so. In fact, there is a hint
of an interstellar absorption line at the velocity of complex~C (see Fig.~2).
Data with higher S/N ratio are needed to resolve this issue.

\section{Conclusions}
\par Forty years after the first attempt (Prata \& Wallerstein 1967), we report
the first successful detection of interstellar \CaII\ H and K absorption from
HVC complex~C. This sets an upper limit on the distance of core CIII (left side
in Fig.~1) of 11.2~kpc. For core CI (right side in Fig.~1) we find a lower limit
of 3.7~kpc, possibly 6.7~kpc. Although the stars are 27\deg\ apart on the sky,
it is still safe to conclude that complex~C is located at Galactocentric radius
$<$14~kpc, and lies high above the Galactic plane ($z$=3--9~kpc). A more precise
determination requires a lower limit for core CIII and an upper limit for CI.
\par Integrating $N$(\HI) across the cloud, we estimate $M$(\HI) as
0.7--6\tdex{6}~\Msun. \Ha\ emission has also been detected (Tufte et al.\ 1998).
We can assume either that the H$^+$ and \HI\ are thorougly mixed or that the
H$^+$ originates in a photoionized skin around the cloud. In either case, the
observed \Ha\ intensity suggests that there is roughly as much ionized as
neutral gas.
\par We can also estimate the mass inflow associated with complex~C, using a
method described in paper~I. Including the neutral and ionized hydrogen, as well
as a 40\% contribution from helium, we derive that complex~C represents about
0.1--0.25~\Msunpyr\ of infalling gas. This is a substantial fraction of the
theoretically required amount of 1~\Msunpyr. Other HVCs may contribute the rest,
but we have not yet determined distances and metallicities for the most likely
candidates.
\par From our results, we conclude that the mystery of the distances to the HVCs
is beginning to be solved. The evidence shows that several HVCs are located in
the upper reaches of the gaseous Galactic Halo and that they contribute
significantly to the inflow of metal-poor gas onto the Galaxy. Once the mass
inflow rate is constrained from observations of a sufficient number of HVCs, the
next step will be to determine their three-dimensional structure, so that we can
use their velocities and galactic location to derive orbits and solve the
outstanding mystery of their ultimate origins.

\bigskip
Acknowledgements
\par B.P.W., D.G.Y, R.W. and T.C.B. acknowledge support from grant AST-06-07154
awarded by the US National Science Foundation. T.C.B. also acknowledges NSF
grants AST 04-06784 and PHY-02-16783; Physics Frontier Center/Joint Institute
for Nuclear Astrophysics (JINA).
\par Some of the data presented were obtained at the W.M. Keck Observatory,
which is operated as a scientific partnership among the California Institute of
Technology, the University of California and the National Aeronautics and Space
Administration. The Observatory was made possible by the generous financial
support of the W.M. Keck Foundation.
\par The William Herschel telescope is operated on the Island of La Palma by the
Isaac Newton group in the Spanish Observatorio del Roque de los Muchachos of the
Instituto de Astrophysica de Canarias.
\par Funding for the SDSS and SDSS-II has been provided by the Alfred P. Sloan
Foundation, the Participating Institutions, the National Science Foundation, the
U.S. Department of Energy, the National Aeronautics and Space Administration,
the Japanese Monbukagakusho, the Max Planck Society, and the Higher Education
Funding Council for England. The SDSS Web Site is http://www.sdss.org/. The SDSS
is managed by the Astrophysical Research Consortium for the Participating
Institutions. The Participating Institutions are the American Museum of Natural
History, Astrophysical Institute Potsdam, University of Basel, University of
Cambridge, Case Western Reserve University, University of Chicago, Drexel
University, Fermilab, the Institute for Advanced Study, the Japan Participation
Group, Johns Hopkins University, the Joint Institute for Nuclear Astrophysics,
the Kavli Institute for Particle Astrophysics and Cosmology, the Korean
Scientist Group, the Chinese Academy of Sciences (LAMOST), Los Alamos National
Laboratory, the Max-Planck-Institute for Astronomy (MPIA), the
Max-Planck-Institute for Astrophysics (MPA), New Mexico State University, Ohio
State University, University of Pittsburgh, University of Portsmouth, Princeton
University, the United States Naval Observatory, and the University of
Washington.

%%%%%%%%%%%%%%%%%%%%%%%%%%%%%%%%%%%%%%%%%%%%%%%%%%%%%%%%%%%%%%%%%%%%%%%%%%%%%%%%
%%%%%%%%%%%%%%%%%%%%%%%%%%%%%%%%%%%%%%%%%%%%%%%%%%%%%%%%%%%%%%%%%%%%%%%%%%%%%%%%

%\newpage\par

\let\ch=\colhead
\def\Texp{T$_{\rm exp}$}
\def\vHVC{v$_{\rm HVC}$}
\def\Wexp{W$_{\rm exp}^g$}
\def\Wobs{W$_{\rm obs}^h$}
\def\kms{kms$^{-1}$}
\begin{deluxetable}{lrrrrrrrrrrrrr}
\rotate
\tabletypesize{\footnotesize} \tabcolsep=2pt \tablenum{1} \tablewidth{0pt} \tablecolumns{7} \tablecaption{Stellar and interstellar data}
\tablehead{%
\colhead{object}         &\ch{$l^a$}  &\ch{$b^a$}  &\ch{dist$^b$} &\ch{v$_*^c$}&\ch{\Texp$^d$}&\ch{S/N$^e$}&\ch{\vHVC$^f$}&\ch{$N$(\HI)$^f$}&\ch{\Wexp}&\ch{\Wexp}&\ch{\Wobs}  &\ch{\Wobs}  \\
                         &            &            &              &           &              &            &              &                 &\ch{(K)}  &\ch{(H)}  &\ch{(K)}    &\ch{(H)}    \\
%                        &\ch{[\deg]} &\ch{[\deg]} &\ch{[kpc]}    &\ch{[\kms]}&\ch{[sec]}    &            &\ch{[\kms]}&\ch{[\dex{18}\cmm2]}&\ch{[\mA]}&\ch{[\mA]}&\ch{[\mA]}  &\ch{[\mA]}  \\
                         &\ch{\deg}   &\ch{\deg}   &\ch{kpc}      &\ch{\kms}  &\ch{sec}      &            &\ch{\kms}    &\ch{\dex{18}\cmm2}&\ch{\mA}  &\ch{\mA}  &\ch{\mA}    &\ch{\mA}    \\
\colhead{(1)}            &\ch{(2)}    &\ch{(3)}    &\ch{(4)}      &\ch{(5)}   &\ch{(6)}      &\ch{(7)})   &\ch{(8)}      &\ch{(9)}         &\ch{(10)} &\ch{(11)} &\ch{(12)}   &\ch{(13)}    }
\startdata
SDSS\,J121611.13+645811.0&   128.94   &   51.74    & 12.5$\pm$0.4 &   $-$306  &     1000     &      8     &    $-$156    &  59.3\e1.1      &   106    &    62    & $<$43      & $<$49      \\
                         &            &            &              &           &              &            &     $-$51    & 127.2\e2.5      &   215    &   161    & 210\e11\e4 & 122\e16\e5 \\
SDSS\,J120404.78+623345.6&   132.12   &   53.71    & 10.9$\pm$0.7 &   $-$389  &     1600     &     20     &    $-$145    &  31.2\e0.8      &    42    &    23    &  42\e 3\e4 &  19\e 6\e3 \\
                         &            &            &              &           &              &            &     $-$48    & 103.1\e1.3      &   186    &   130    & --         & 113\e15\e3 \\
%DSS\,J120557.16+625251.6&   131.61   &   53.47    &  9.3$\pm$0.9 &   $-$164  &      600     &     24     &    $-$147    &  38.6\e1.3      &    58    &    32    & --         & --         \\
%                        &            &            &              &           &              &            &     $-$48    & 108.6\e2.0      &   194    &   137    & 239\e10\e8 & --         \\
BS\,16079-0017           &    91.05   &   46.60    &  8.1$\pm$2.9 &   $-$208  &     1800     &     20$^k$ &    $-$139    &  32.5\e1.2      &    45    &    24    & $<$16      & --         \\
BS\,16034-0114$^i$       &    89.39   &   45.07    &  3.8$\pm$0.3 &       40  &     3600     &     16     &    $-$123    &  71.6\e1.3      &   132    &    82    & $<$12      & $<$21      \\
BS\,16079-0015$^j$       &    90.69   &   46.46    &  2.0$\pm$0.2 &   $-$234  &     1800     &     62$^l$ &    $-$134    &  50.1\e3.0      &    85    &    48    &  $<$8      & --         \\

%S\,16034.0002           &    91.60   &   48.86    &  8.4$\pm$1.0 &    $-$19  &     1800     &     16     &    $-$119    &  35.1\e2.0      &    50    &    27    & $<$22      & $<$39      \\
%S\,16086.0123$^m$       &    92.14   &   47.40    &  3.7$\pm$0.5 &    $-$60  &     1800     &      7     &    $-$139    &  44.6\e1.6      &    72    &    40    & $<$24      & $<$22      \\
\enddata
\tablecomments{%
a: Columns~2 and 3 give the Galactic longitude and latitude of the stars;
b: Column~4 is the distance as determined by Wilhelm et al.\ (2008, in preparation);
c: Column~5 gives the stellar velocity (relative to the LSR), as measured from \CaII\ and \FeI\ lines.
d: Column~6 is the exposure time with the WHT (BS stars) or Keck (SDSS stars);
e: Column~7 gives the S/N ratio in \CaII\ K near the HVC velocity;
f: Columns~8 and 9 give the velocity (relative to the LSR) and \HI\ column density of complex~C in the
direction of the star. Values are based on the LAB survey (Kalberla et al.\
2005) for SDSS stars, and on our Effelsberg data (Wakker et al.\ 2001) for BS
stars; 
g: Columns~10 and 11 give the expected K and H equivalent widths, using the
relation log($N$(\CaII)/$N$(\HI)) = $-$7.76$-$0.78\,(log($N$(\HI))$-$19.5) found by
Wakker \& Mathis (2000). Note that the uncertainty in $N$(\HI) due to the
large radio beamsize produces an uncertainty of about 20~\mA\ in these prediction.
h: Columns~12 and 13 show the observed K and H equivalent widths or 3$\sigma$
upper limits;
i: alternative names 2MASS\,J160007.91+575125.4 and BS16079-0065;
j: alternative name 2MASS\,J154747.53+580646.4;
k: there are too many stellar lines to calculate an \CaII~H error;
l: the error used for calculating the S/N includes variations due to many weak stellar lines.}
%m: alternative name 2MASS\,J153753.36+583317.6;
\end{deluxetable}

%fiddle: size angle hscale vscale hoff voff
\begin{figure}\plotfiddle{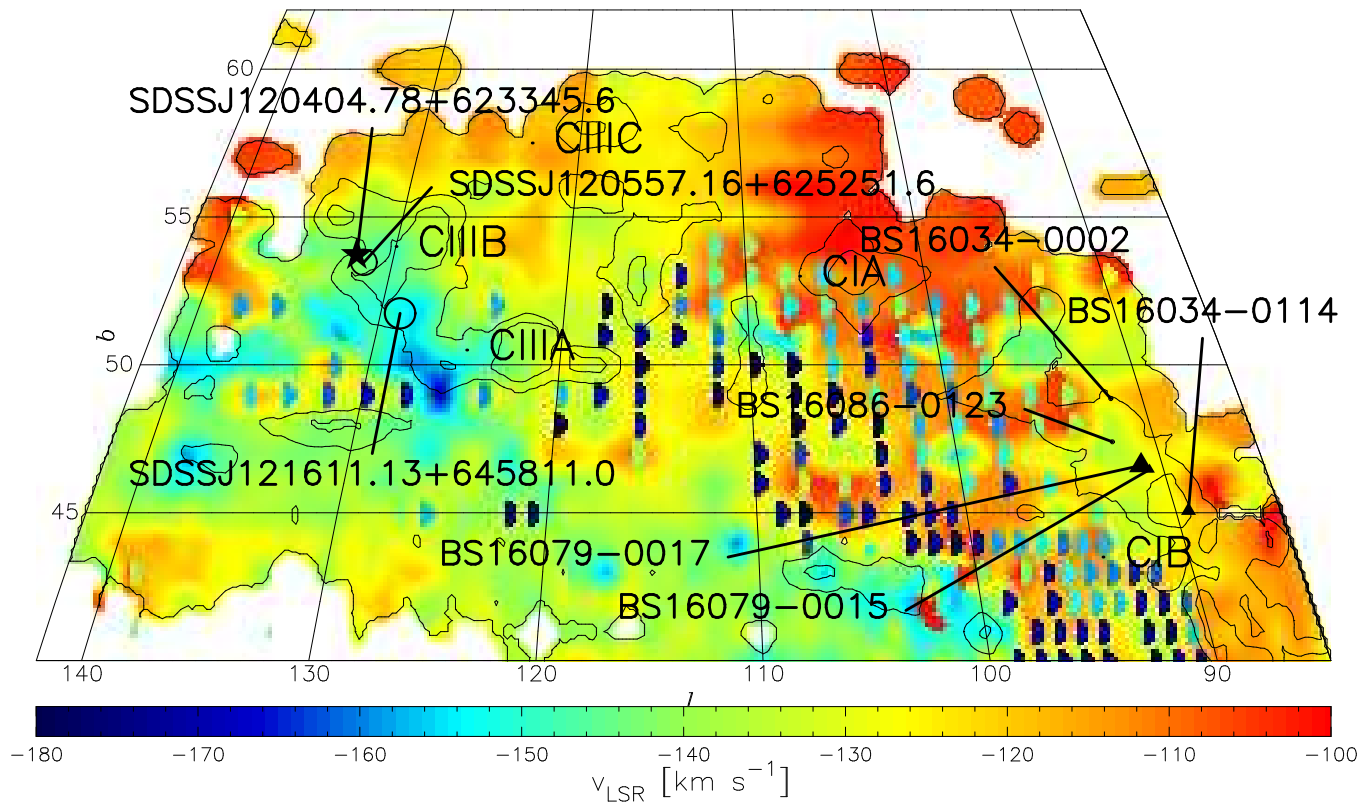}{0.1in}{270}{300}{510}{0}{0}\figurenum{1} % APJMS
\caption{%
\HI\ map of complex~C, based on the data of Hulsbosch \& Wakker (1988). Colors
represent LSR velocities, as coded in the wedge. Half circles show positions
with multiple \HI\ components. Contour levels are at brightness temperatures of
0.05, 0.25, 0.5 and 1~K. The positions of the stars discussed in this paper are
shown by the symbols -- closed stars for the detections, closed triangles for
the significant non-detections, open circles for non-significant non-detections.
The symbol diameters are proportional to the stellar distances. Several cores
inside complex~C are labeled.
}\end{figure}

\begin{figure}\plotfiddle{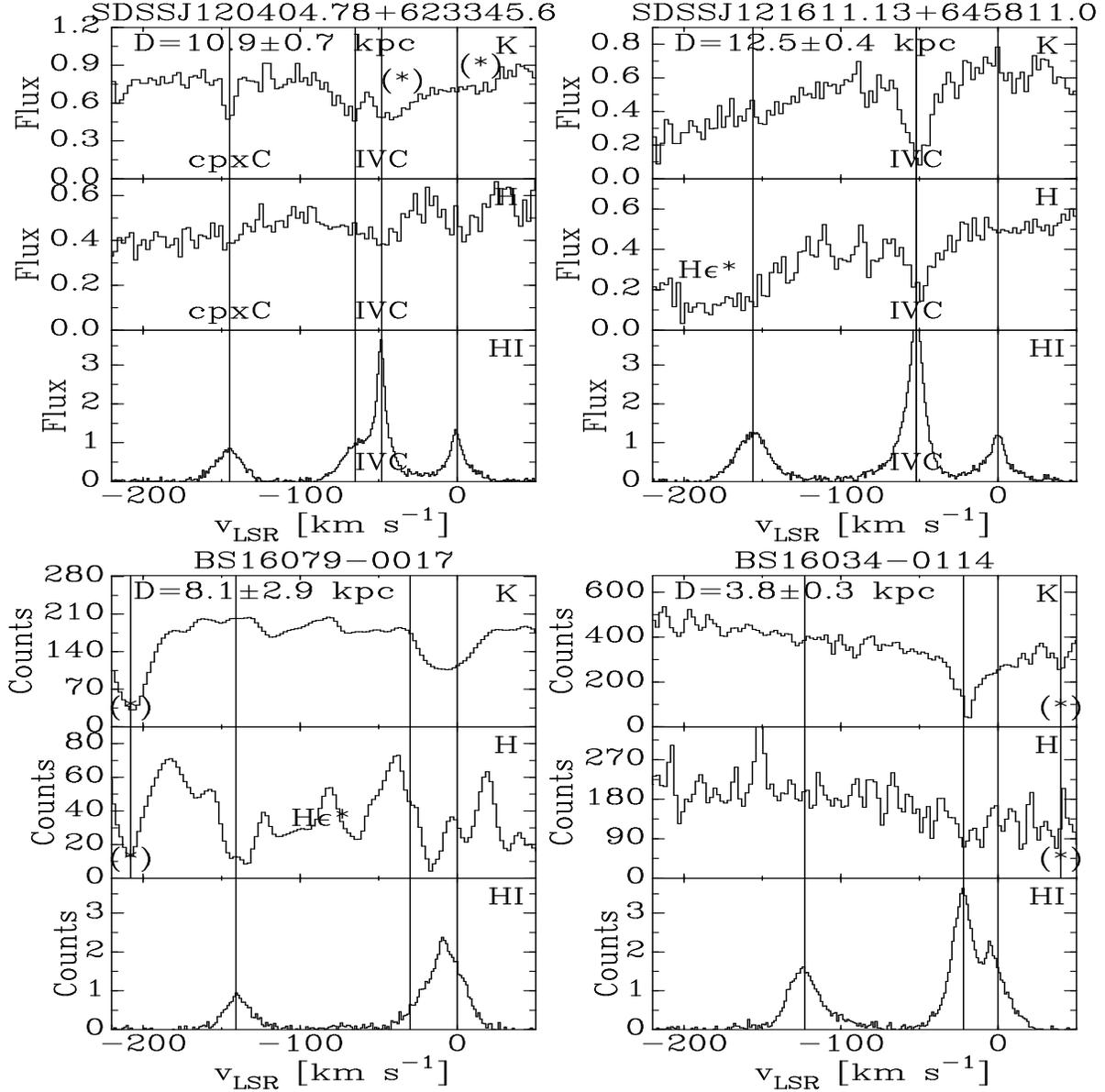}{0in}{0}{450}{450}{0}{0}\figurenum{2} % APJMS
%\begin{figure}\plotone{f2.eps}\figurenum{2} % APJMS
\caption{%
Spectra near \CaII\ K, H and \HI\ for the four most significant stars in our
sample. Note that the Keck data are flux-calibrated (units \dex{-15}
erg\,\cmm2\,s$^{-1}$\,\AA$^{-1}$), while the WHT data are not. Vertical lines
are placed at the velocity of the low-, intermediate- and high-velocity \HI\
emission components, while detected absorption lines are labeled ``cpxC'' and
``IVC''. The locations of stellar \CaII\ and \FeII\ lines are shown by the (*).
Note that near the H line toward BS\,16079-0017 there are many unidentified
stellar lines.
}\end{figure}

\begin{figure}\plotfiddle{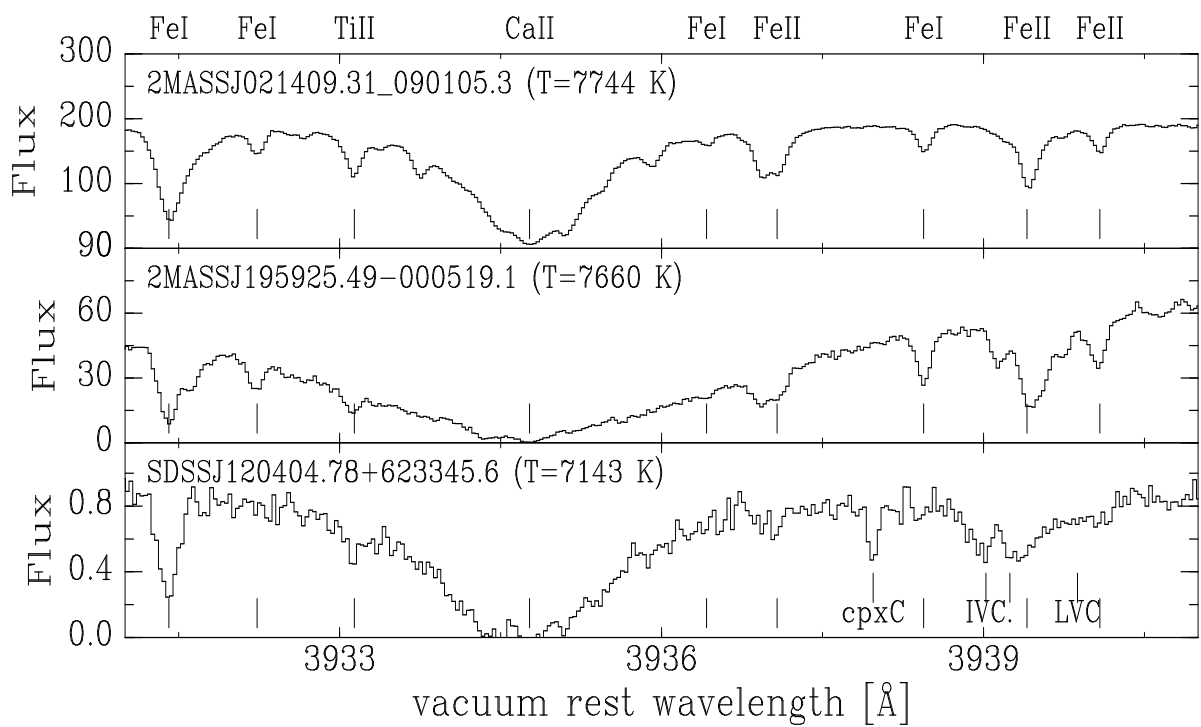}{0in}{0}{450}{270}{0}{0}\figurenum{3} % APJMS
%\begin{figure}\plotone{f3.eps}\figurenum{3} % APJMS
\caption{%
Spectra of three stars, shifted to the stellar reference frame. The top two
(2MASS) stars were observed with the VLT by Wakker et al.\ (2007; paper~I). Most
of the stellar lines can be identified as \FeI\ or \FeII\ absorption. Some
features have not yet been identified. Clearly, no stellar line is expected at
the wavelength of the complex~C \CaII~K absorption. The two-component
intermediate-velocity and low-velocity \CaII~K absorption in the direction of
SDSS\,J120404.78+623345.6 blends with stellar \FeII\ absorption.
}\end{figure}

\end{document}